%% file: main.tex
\documentclass[9pt, conference]{IEEEtran}
\usepackage{cite}
\usepackage{amsmath,amssymb,amsfonts}
\usepackage{algorithm}
\usepackage{algpseudocode}
\usepackage{multirow}
\usepackage{graphicx}
\usepackage{subcaption}
\usepackage{textcomp}
\usepackage{url}
\usepackage[hidelinks]{hyperref}
\usepackage[dvipsnames]{xcolor}
\usepackage{booktabs}
\usepackage{fancyhdr}

\def\BibTeX{{\rm B\kern-.05em{\sc i\kern-.025em b}\kern-.08em
    T\kern-.1667em\lower.7ex\hbox{E}\kern-.125emX}}

\begin{document}

\title{High-Resolution Speech Restoration with Latent Diffusion Model}










\author{
\begin{tabular}{@{}c@{}}

Tushar Dhyani$^{1,2}$
\qquad Florian Lux$^{2}$
\qquad Michele Mancusi$^{1}$
\qquad Giorgio Fabbro$^{1}$
\qquad Fritz Hohl$^{1}$
\qquad Ngoc Thang Vu$^{2}$
\end{tabular}
\IEEEauthorblockA{ \\
\\
$^1$ Sony Europe B.V., Stuttgart, Germany \quad $^2$ University of Stuttgart, Germany \\
\hspace{1.5cm} $^1$ firstname.lastname@sony.com \hspace{0.25cm} \quad $^2$ firstname.lastname@ims.uni-stuttgart.de \\
}
}

\maketitle

\begin{abstract}

Traditional speech enhancement methods often oversimplify the task of restoration by focusing on a single type of distortion. Generative models that handle multiple distortions frequently struggle with phone reconstruction and high-frequency harmonics, leading to breathing and gasping artifacts that reduce the intelligibility of reconstructed speech. These models are also computationally demanding, and many solutions are restricted to producing outputs in the wide-band frequency range, which limits their suitability for professional applications.
To address these challenges, we propose Hi-ResLDM, a novel generative model based on latent diffusion designed to remove multiple distortions and restore speech recordings to studio quality at a full-band sampling rate of 48kHz. Benchmarked against state-of-the-art methods that leverage GAN and Conditional Flow Matching (CFM) components, Hi-ResLDM demonstrates superior performance in regenerating high-frequency-band details. Hi-ResLDM not only excels in non-instrusive metrics but is also consistently preferred in human evaluation and performs competitively on intrusive evaluations, making it ideal for high-resolution speech restoration.

\end{abstract}


\input{sections/intro}
\input{sections/methods}

\input{sections/experiment}
\input{sections/results}

\input{sections/conclusion}

\clearpage
\bibliographystyle{IEEEbib}
\bibliography{main}


\end{document}

%% file: sections/intro.tex
\section{Introduction}

Generative speech restoration \cite{liu_voicefixer_2022, resemble_enhance, gesper, koizumi_miipher_2023} has emerged as a robust solution, addressing the limitations of traditional \cite{ephraim1984speech, PALIWAL2012282} and deep neural network (DNN) based discriminative approaches \cite{defossez2020real, zhao_dereverb_2020, tgerkman_dereverb_2014}.
Inspired by speech processing in the human brain \cite{firststage, secondstage}, these methods use a two-stage mechanism \cite{liu_voicefixer_2022, resemble_enhance, gesper}, consistently outperforming single-stage models \cite{defossez2020real, zhao_dereverb_2020}. 
They not only overcome the limitations of compartmentalized models but also handle multiple distortions simultaneously \cite{serra_universal_2022}. Another classifier-based approach \cite{rice_general_audio_2023} shows potential for general audio effect removal but lacks the ability to generalize to simultaneous distortions in speech recordings, suggesting opportunities for enhancing existing methods.

DNN-based discriminative methods extract information from noisy speech, while generative methods, conditioned on noisy inputs, mimic the underlying speech distribution. In a low signal-to-noise ratio (SNR) scenario, generative models may approximate the distribution incorrectly, producing erroneous phones, reducing intelligibility, and causing phonetic confusion. These incoherent generated phones are called breathing and gasping artifacts \cite{de2023behavior, lemercier_storm_2023}. Discriminative methods, however, often eliminate low-energy speech regions from low-SNR recordings, a limitation not seen in generative models \cite{lemercier2023analysing}.

\begin{figure}[!t]
  \centering
  \begin{subfigure}[b]{0.35\linewidth}
  \centering
    \includegraphics[width=\linewidth]{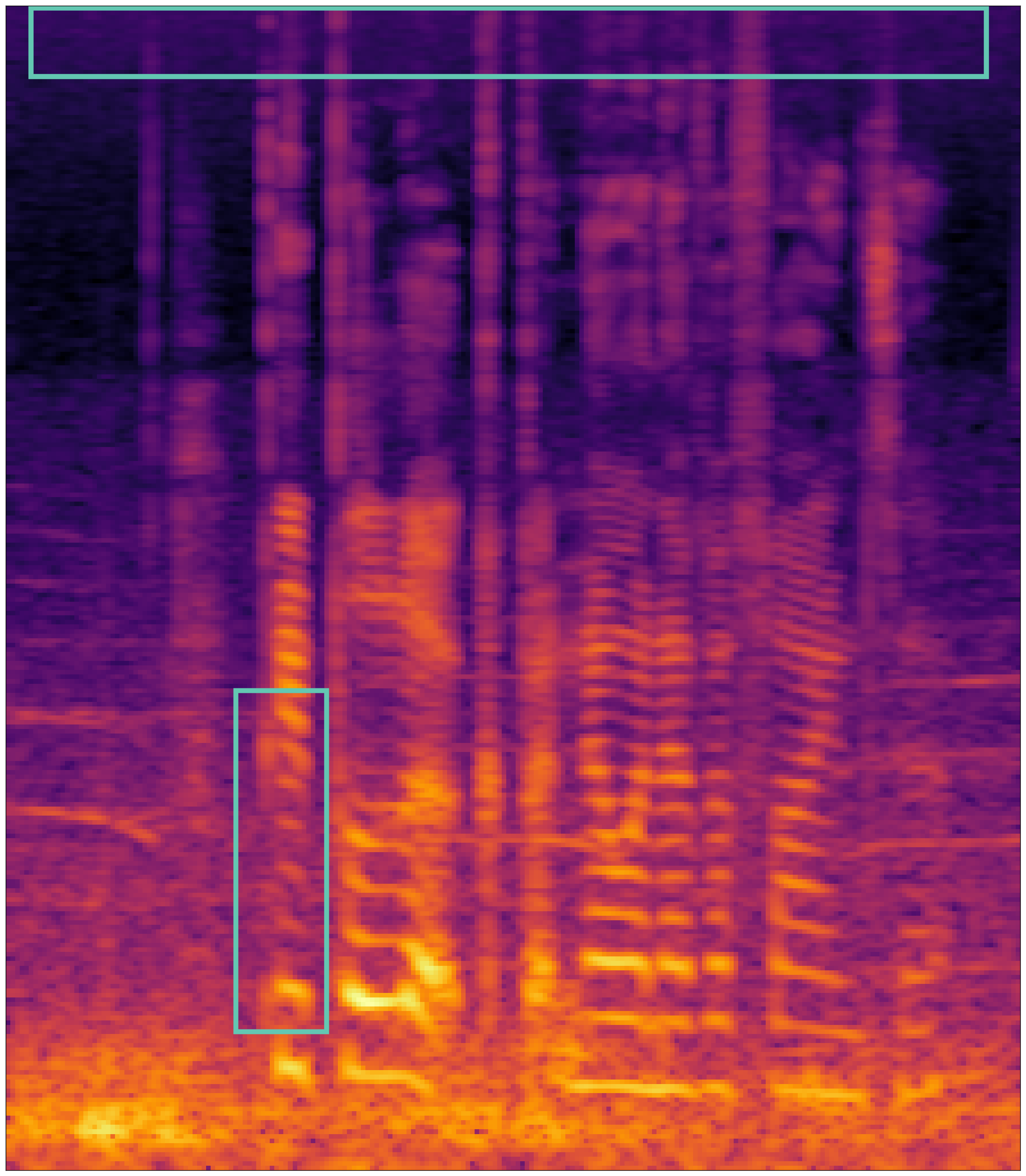}
    \caption{Noisy spectrogram}
    \label{fig:target}
    
  \end{subfigure}
 \hspace{0.5cm}
  \begin{subfigure}[b]{0.35\linewidth}
  \centering
    \includegraphics[width=\linewidth]{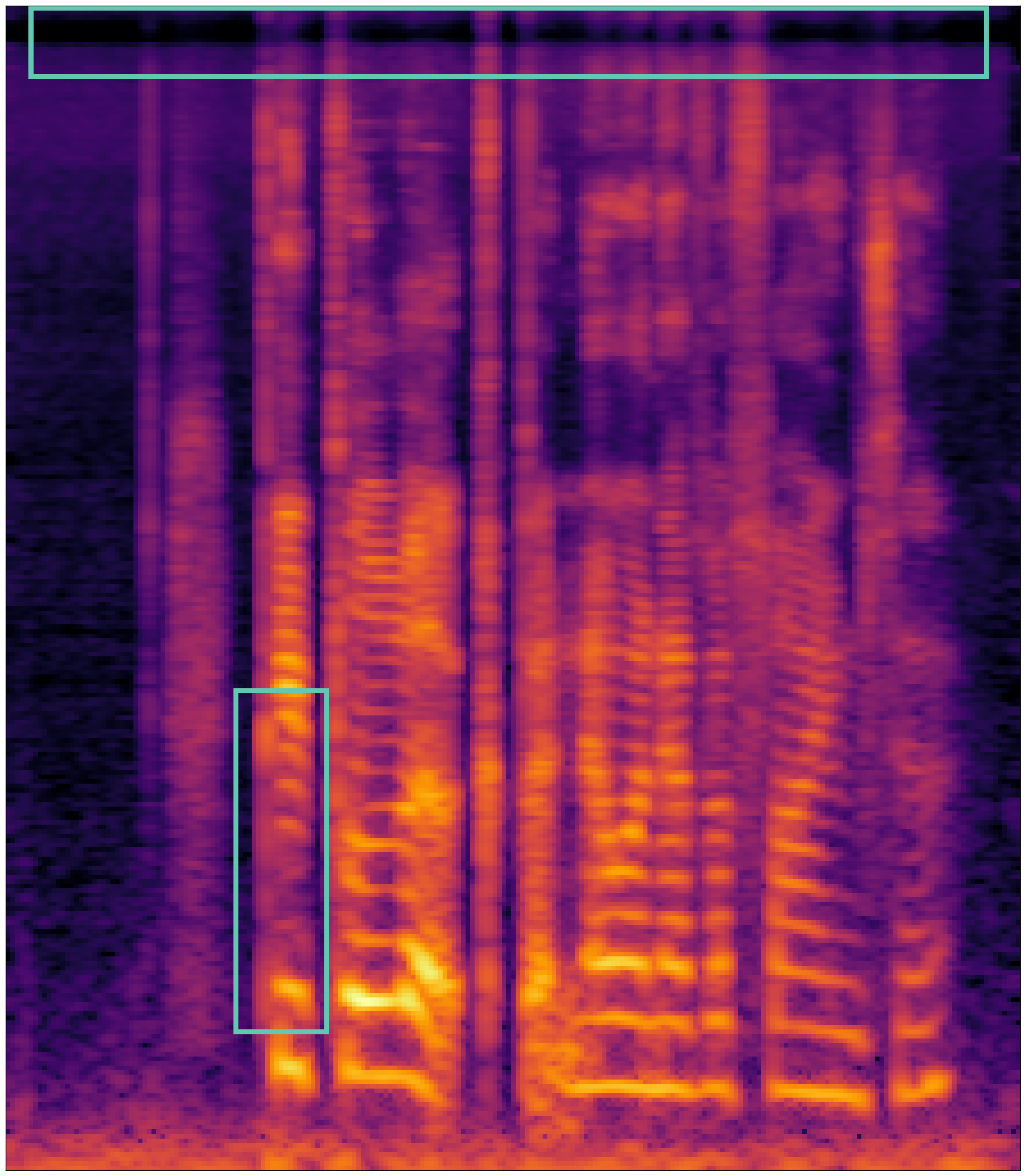}
    \caption{VoiceFixer}
    \label{fig:voicefixer_output}
  \end{subfigure}
  \medskip

  \begin{subfigure}[b]{0.35\linewidth}
    \centering
    \includegraphics[width=\linewidth]{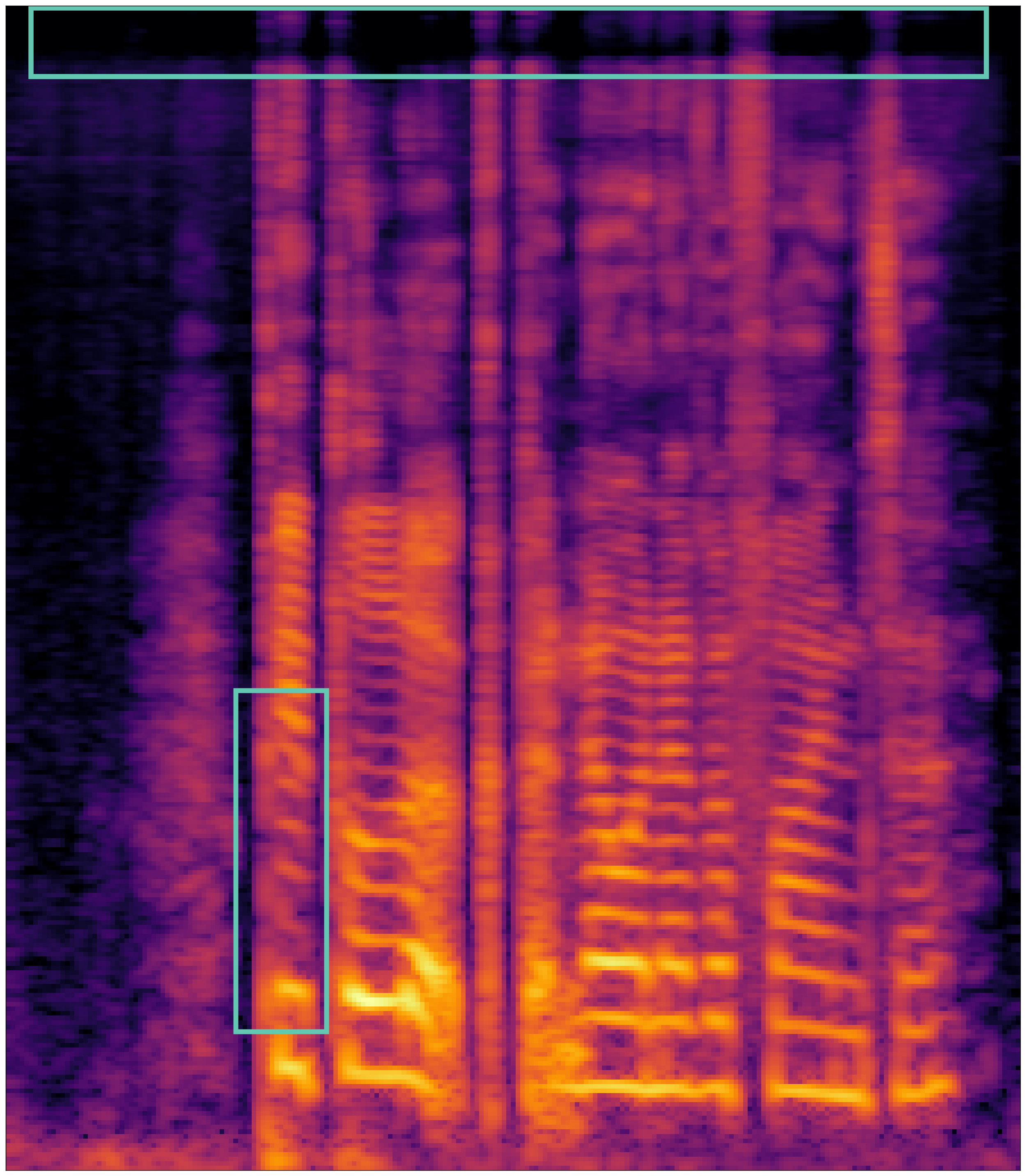}
    \caption{Resemble Enhance}
    \label{fig:resem_output}
  \end{subfigure}
   \hspace{0.5cm}
  \begin{subfigure}[b]{0.35\linewidth}
    \centering
    \includegraphics[width=\linewidth]{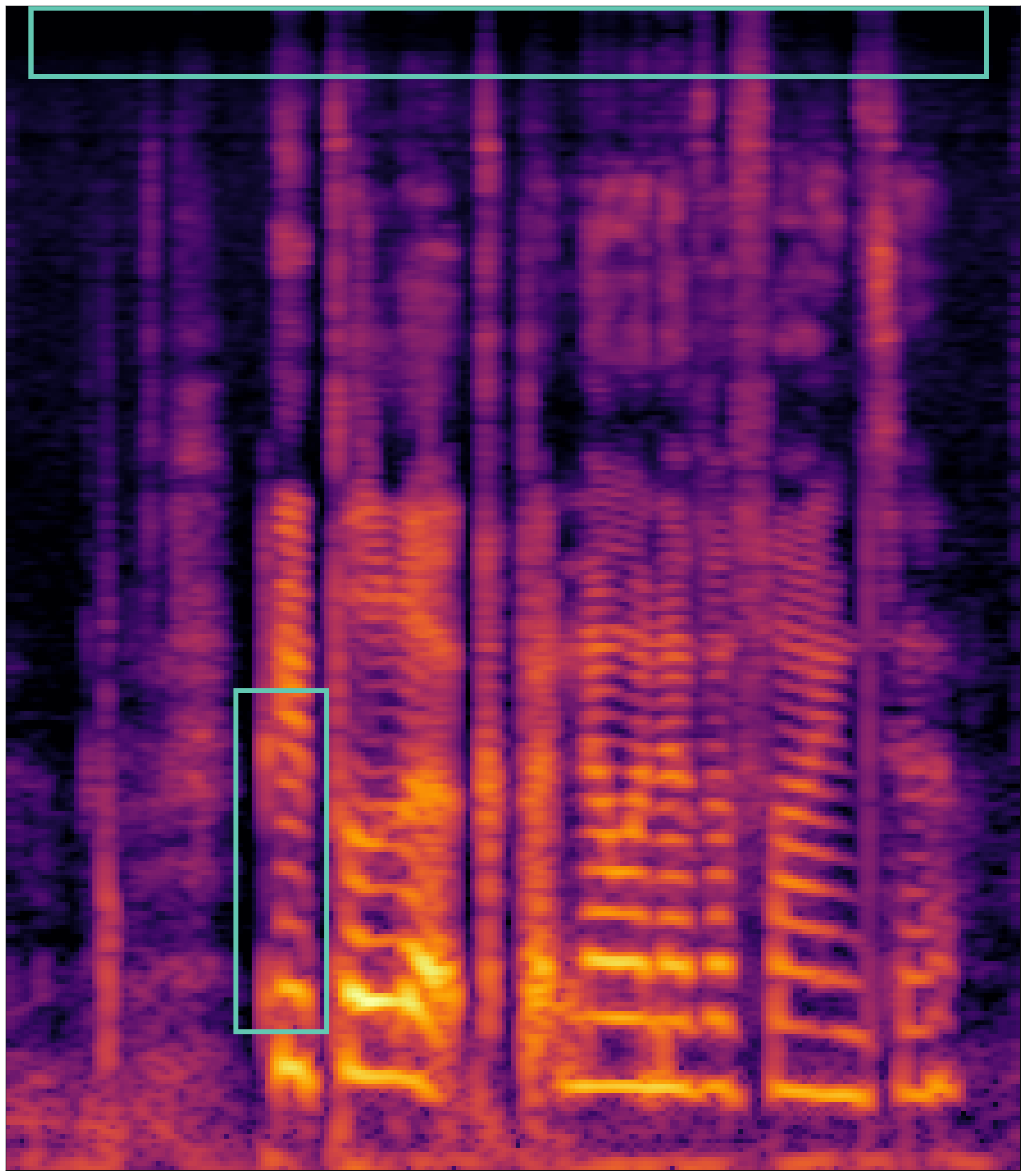}
    \caption{Hi-ResLDM (LDM)}
    \label{fig:our_ldm_output}
  \end{subfigure}
  
  \caption{Mel-spectrograms of a restored speech signal. The green highlighted rectangles emphasize the sections where the harmonic structure generated by Hi-ResLDM is prominently better compared to Voicefixer \cite{liu_voicefixer_2022} and Resemble Enhance \cite{resemble_enhance}. }
  \label{fig:all_images}
\end{figure}

Single-stage restoration methods often overfit during the filtering process \cite{sulun_single_overfitting_2021}. To address the issue of generalizability, two-stage approaches have been effective for generative speech restoration \cite{liu_voicefixer_2022, gesper}. These methods typically involve an initial generative enhancement stage followed by a vocoding step for the mel-spectrogram inversion, often utilizing generative adversarial networks (GANs) \cite{goodfellow2014generativeadversarialnetworks}. However, a key limitation is that the first stage handles most of the enhancement, leaving the second stage to have no significant impact on restoration capability. An alternative solution involves combining a discriminative method with a generative model \cite{lemercier_storm_2023}. The primary advantage of this approach is that the generative model is tasked with filling in the missing information rather than reconstructing the entire signal, thereby addressing the limitations of the discriminative stage. However, this strategy underutilizes the generative model's full potential, as it is confined to refining residual details instead of performing comprehensive signal regeneration.

\begin{figure*}
    \centering
    \includegraphics[width=1\linewidth]{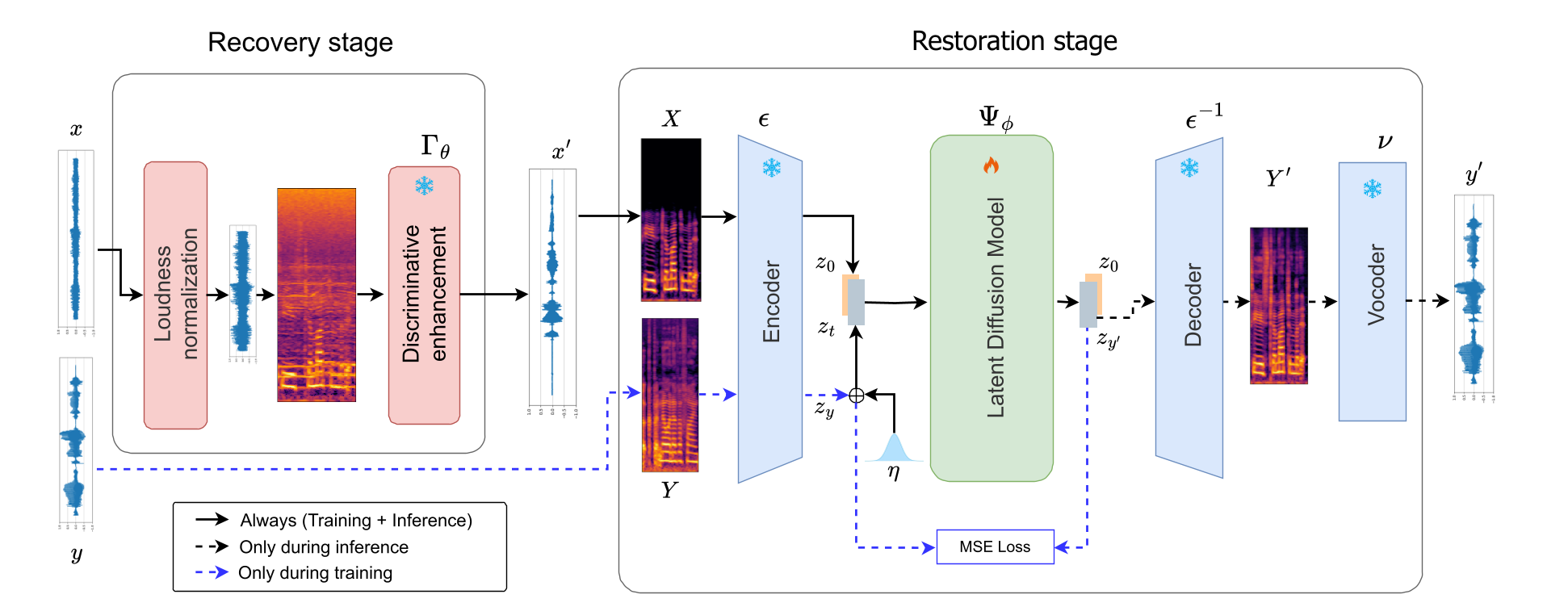}
    \caption{A high-level overview of the Hi-ResLDM model, illustrating the components of the proposed two-stage approach. The black arrow connects components used during both training and inference. The blue dashed line connects the training components only, and the black dashed line connects components only used during inference. }
    \label{fig:Overview-figure}
\end{figure*}

To address the stated limitations of existing two-stage approaches, we propose:
\begin{itemize}
    \item Hi-ResLDM\footnote{\url{https://github.com/sony/Hi-ResLDM}}, a novel two-stage framework, shown in Fig. \ref{fig:Overview-figure}, for robust speech restoration capable of handling multiple distortions simultaneously. Our approach combines discriminative and diffusion-based generative methods operating within the latent space \cite{rombach2022ldm} of an autoencoder. Hi-ResLDM is specifically designed for restoring speech to resemble studio-quality, full-band recordings, typically sampled at $48\text{kHz}$.
    \item We test restoration frameworks for iterative refinement, a popular technique in image restoration \cite{saharia2021image}, and demonstrate that Hi-ResLDM returns consistent reconstructions across multiple refinements.
    \item We compare Hi-ResLDM with popular publicly available models, showing that it outperforms GAN-based Voicefixer (VF) \cite{liu_voicefixer_2022} and conditional flow matching (CFM) \cite{lipman2022cfm} based Resemble Enhance (RE) \cite{resemble_enhance} in general speech restoration. However, this improvement comes with increased inference time, offering better harmonic generation and balanced low-energy restoration, as illustrated in Fig. \ref{fig:all_images}.
\end{itemize}

%% file: sections/methods.tex
\section{Method}

\begin{table*}[t]
    \centering
    \begin{tabular}{l c c c c c c c c }
        \toprule
        \multirow{2}{*}[-2pt]{\textbf{Method}} & \multicolumn{4}{c}{\textbf{Valentini dataset}} & \multicolumn{4}{c}{\textbf{Internal testset}} \\
                          \cmidrule{2-9} 
         & \textbf{eSTOI}($\uparrow$)  & \textbf{DNSMOS}($\uparrow$) & \textbf{NISQA}($\uparrow$) & \textbf{WER}($\downarrow$) & \textbf{eSTOI}($\uparrow$)  & \textbf{DNSMOS}($\uparrow$) & \textbf{NISQA}($\uparrow$) & \textbf{WER}($\downarrow$)\\
        \midrule
        Noisy           & $0.78 \pm 0.14$                           & $3.05 \pm 0.38$            &  $2.99 \pm 0.73$ &  $0.08 \pm 0.24$ & $0.62 \pm 0.18$                          & $2.81 \pm 0.38$                           &  $2.23 \pm 0.84$ &  $0.10 \pm 0.26$ \\ 
        Clean           & -                           & $3.53 \pm 0.27$           & $4.62 \pm 0.29$ & - & - & $3.53 \pm 0.27$           & $4.62 \pm 0.29$ & -\\
        \midrule
        
        VoiceFixer \cite{liu_voicefixer_2022}      & $0.75 \pm 0.07$           & $3.41 \pm 0.29$           & $4.43 \pm 0.39$ & $0.09 \pm 0.24$ & $0.72 \pm 0.09$           & $3.37 \pm 0.28$           & $4.07 \pm 0.46$ & $0.13 \pm 0.29$\\ 
        
        Resemble Enhance \cite{resemble_enhance} & $0.79 \pm 0.08$          & $\textbf{3.56} \pm \textbf{0.26}$           & $4.51 \pm 0.35$ & $0.07 \pm 0.19$ & $0.74 \pm 0.11$          & $\textbf{3.61} \pm \textbf{0.26}$           & $4.30 \pm 0.47$ & $0.14 \pm 0.27$\\ 
        
        Hi-ResLDM (Recv.)            & $\textbf{0.84} \pm \textbf{0.09}$ & $3.39 \pm 0.21$    & $4.37 \pm 0.47$ & $\textbf{0.05} \pm \textbf{0.19}$ & $0.72 \pm 0.12$ & $3.42 \pm 0.29$   & $4.39 \pm 0.61$ & $\textbf{0.11} \pm \textbf{0.28}$\\  
        Hi-ResLDM    & $0.82 \pm 0.09$ & $3.48 \pm 0.27$    & $\textbf{4.54} \pm \textbf{0.54}$ & $\textbf{0.05} \pm \textbf{0.15}$ & $\textbf{0.76} \pm \textbf{0.12}$ & $3.46 \pm 0.29$   & $\textbf{4.42} \pm \textbf{0.54}$ & $0.12 \pm 0.26$\\  
        \bottomrule 

    \end{tabular}
    \vspace{0.1cm}
    \caption{Results of restoration approaches on noisy and clean Valentini dataset \cite{valentini2017noisy} and our internal test set using intrusive and non-intrusive evaluation methods. In the table, Recv. represents the results of only the recovery stage operating at $16$kHz.}
    \label{res:tab:overall results}
\end{table*}

Given a monaural noisy signal $x(t) \in \mathbb{R}^T$ with sample rate $r$, it can be expressed as a mixture of clean speech sample $y(t) \in \mathbb{R}^T$, and some additive noise $n(t) \in \mathbb{R}^T$. Occasionally, the signal can be an impulse response $h(t) \in \mathbb{R}^{t_{60}}$, and various distortion functions such as clipping, low-pass filter, and resampling collectively represented by $d(t)$. So, it can be formulated as $x(t) = d((y(t) + n(t)) \ast h(t))$, where $\ast$ denotes the convolution operation between the noisy input $y(t) + n(t)$ and the impulse response $h(t)$. For the sake of simplicity, from now on, we will indicate all functions of time $f(t)$ as $f$. 

To simplify the process of distortion inversion, we adopted a two-stage approach that has demonstrated state-of-the-art performance in speech inversion tasks \cite{liu_voicefixer_2022, gesper}. In the first stage (recovery stage), the distorted input $x$ is fed into a model $\Gamma_\theta: x \xrightarrow{}x^\prime$ that removes additive distortions and gives a clean, intermediate downsampled estimate $x^\prime$. In the second stage (restoration stage), a model $\Psi_{\phi}: x^\prime \xrightarrow{} y^\prime$ further processes $x^\prime$ to regenerate the final clean speech estimate $y^\prime$. 

\subsection{Recovery stage}

The goal of the first stage is to increase SNR by removing the noise $n$ from $x$.
To achieve this, $\Gamma_\theta$ is decomposed into two sub-steps: loudness normalization and discriminative enhancement. While a neural network can learn loudness normalization as a linear operation, separating this step offers finer control over the preprocessing of input signals. Isolating loudness normalization from the enhancement step amplifies the target energies in $x$, enabling the discriminative enhancement stage to focus more effectively on $x^\prime$.
Thus, the primary task of $\Gamma_\theta$ is removing $n$ from $x$. In the input signal represented as $x = d((y + n) \ast h)$, the convolution operation is applied to the noisy part $y + n$. Due to the linear nature of convolutions, we can simplify it as $x = d(y \ast h + n \ast h)$ and as $n \ast h \approx n$, we re-write the equation as $x = d(y \ast h + n)$. In this formulation, the discriminative enhancement stage is responsible for removing additive degradation from the input, allowing the second stage to focus on generating clean speech without being affected by irrelevant additive information. Since this stage recovers the signal from additive distortions, we refer to the first stage as the \textit{recovery stage}.

We first resample $x$ from $r$ of $16\text{kHz}$ followed by normalization using PyLoudNorm \cite{steinmetz2021pyloudnorm}. 
To avoid clipping in low SNR $x$ due to loudness normalization, we set the loudness to $-20 \text{LUFS}$ (Loudness Units Full Scale) rather than the ITU 1770 (International Telecommunication Union) standard of $-14 \text{LUFS}$. 
The normalized input was first converted into a time-frequency domain using a window size of $32$ms and hop length of $8$ms following the preprocessing steps from \textit{StoRM} \cite{lemercier_storm_2023}. For enhancement, we used the discriminative \textit{NCSN++M} from \textit{StoRM}, chosen for its efficiency and fast inference while maintaining intelligibility. We retrained the network on our training dataset \ref{sec:speech_dataset_train}. The results of our \textit{NCSN++M} model are presented in table \ref{res:tab:overall results}. The output is converted back to the time domain by inverting the complex spectrogram. 

\subsection{Restoration stage}
In this stage, called \textit{restoration stage}, we employ a latent diffusion model (LDM) $\Psi_{\phi}$ \cite{rombach2022ldm, liu2023audioldm, liu2023audioldm2} for the generation of the high-fidelity clean speech signal. Compared to other generative methods, LDM is more stable to train and has the capability of preserving high-frequency details with less computational resources than its spatial counterpart \cite{rombach2022ldm}. First, to train the LDM, $x^\prime$ is converted from the time domain to a time-frequency domain $X$. A pre-trained general audio autoencoder is used to convert $X$ to perceptually equivalent, lower-dimensional latent representation $z_0$ using only the encoder $\epsilon$. $z_0$ is used as conditioning information to train the denoising diffusion probabilistic model (DDPM) \cite{nichol2021improvedddpm}. Here, DDPM learns to estimate the latent representations $z_{y^\prime}$ from latents of target speech $z_y$ given $z_t = \sqrt{\bar{\alpha}_t} z_y + \sqrt{1 - \bar{\alpha}_t} \boldsymbol{\eta}, \quad \boldsymbol{\eta} \sim \mathcal{N} (\mathbf{0}, \mathbf{I})\,$ where $t \in [0, T]$ is the time and $\bar{\alpha}_t$ is defined as in \cite{ho2020ddpm}. The decoder $\epsilon^{-1}$ of the autoencoder converts the estimation $z_{y^\prime}$ to the reconstructed time-frequency representation $Y^\prime$ which is further mapped to the estimation $y^\prime$ in the time domain using an inverter function $\nu$. 

In particular, we used AudioMAE \cite{huang2022MAE} as the autoencoder, a U-Net for the model $\Psi_{\phi}$, and the HiFi-GAN \cite{kong2020hifi} vocoder as the function $\nu$. The output of $\Gamma_\theta$, sampled at 16kHz, is first upsampled to $48$kHz, converted into mel-spectrogram $X$, and then fed into $\epsilon$, finally obtaining its latent representation $z_0$, which will be used as conditioning. Similarly, we also map the clean speech $y$ to its latent representation $z_y$. Before performing DDPM, $z_y$ and $z_0$ are concatenated. We parameterize our approach as in \cite{liu2023audioldm2}, and thus, the training objective for our diffusion model becomes as shown in equation \ref{eq:ldm_objective}. In the end, the reconstructed mel-spectrogram $Y^\prime$ now represents a distortion-less restored clean-speech signal. 

\begin{equation}
\label{eq:ldm_objective}
\min_\phi \mathbb{E}_{z_y, z_0, z_t, t}\left[\Vert z_{y} -  \Psi_{\phi}(z_t, z_0, t) \Vert^2_2\right]\,,
\end{equation}

%% file: sections/experiment.tex
\section{Experimental Setup}

\subsection{Data}
\subsubsection{Training dataset} 
\label{sec:speech_dataset_train}
To achieve our goal of restoring distorted speech to clean, full-band audio, we require a dataset of high-quality, distortion-free speech sampled at $48\text{kHz}$. We trained our individual staged using several open-source datasets, including VCTK \cite{veaux2017cstr, hqvctk}, and a few high-quality open-source speech datasets from OpenSLR\footnote{\url{https://openslr.org/}} \cite{kjartansson-etal-tts-sltu2018_hq_indic, demirsahin-etal-2020-open_hq_british, guevara-rukoz-etal-2020-crowdsourcing_hqlatin, gutkin-et-al-hq_yoruba2020}. Additionally, we incorporated internal datasets, similar to the open-source ones, consisting of single-speaker monaural monologues. To further ensure the cleanliness of the speech recordings, we evaluated each file using the non-intrusive NISQA \cite{mittag2021nisqa} metric and discarded all files having a mean opinion score (MOS) below $4$. A threshold value of $4$ was selected based on sufficient perceived quality. Our curated dataset was resampled to $48$kHz and comprised a total of $1250$ hours of clean recordings. We normalize the loudness of all files to $-20$ LUFS. These diverse datasets across multiple languages provide a versatile distribution of speech characteristics, balanced gender ratios, varying tones, and different accents. As another pre-processing step, silence was removed from all datasets for training, and each recording was split into chunks of $5.12$ seconds. For augmenting distortions wherever necessary, reverberations were applied equally from simulated and internal Impulse response (IR) datasets, keeping their reverberation times (T60) below $1s$. For additive noise, we used FSD50K \cite{fonseca2022FSD50K}. Noise and reverberation samples were equally divided into 80\% and 20\% for training and evaluation split, respectively.


\subsubsection{Evaluation dataset}
\label{sec:speech_data_eval}
We evaluated all methods on speech enhancement and reverberation tasks using the VCTK test split \cite{liu_voicefixer_2022} and an internal set created per \ref{sec:speech_dataset_train}. Additive noise was introduced keeping the SNR values between $-5$dB and $10$dB, and T60 was maintained between $0.1$ and $0.5s$. Care was taken to keep evaluation speech samples and distortions separate from the training set to avoid data leakage. Furthermore, artifacts from OPUS and Vorbis codecs were introduced in 50\% of the data.

\subsection{Evaluation protocol}

Hi-ResLDM, as a generative method, lacks a one-to-one mapping with the target signal, necessitating the use of non-intrusive evaluation metrics. Intrusive metrics, which require sample alignment between the predicted and target signals, are less effective for generative techniques. However, to evaluate our first stage, we used two intrusive metrics, extended STOI \cite{taal_estoi_2010}, which operates at $10\text{kHz}$, and structural similarity (SSIM) at different sampling rates. Despite eSTOI being intrusive, offers a basis for assessing speech intelligibility, whereas SSIM provides a comparison of phonemic structural fidelity of the generated speech signal. 
Although contrastive evaluation methods existed \cite{ciranni_cocola_2024}, a speech-specific measure was unavailable at the time of writing, so we excluded them.

For the primary evaluation, we relied on two established non-intrusive speech evaluation metrics: DNSMOS \cite{9414878} and NISQA \cite{mittag2021nisqa}. DNSMOS, operating on $16$kHz audio, estimates overall perceptual quality, while NISQA, using smaller overlapping segments and two neural networks, penalizes artifacts such as pre-phonetic breathing and extended fricatives but emphasizes accurate phoneme generation.

In addition to intrusive and non-intrusive metrics, a key objective of Hi-ResLDM is to preserve the fidelity of generated phonemes. To assess phonemic confusion, we transcribed the output using the open-source Whisper-large model \cite{radford_whisper_2023} and calculated the Word Error Rate (WER) by comparing the transcriptions with human-annotated transcripts of our evaluation dataset. We evaluate speaker consistency using speaker recognition cosine similarity (SR-CS), which measures the distance between embeddings of target and restored files generated by speaker-verification model \cite{desplanques20_ecapa_interspeech}. 

To evaluate Hi-ResLDM’s real-world performance, we conducted a subjective listening test in which 20 volunteer audio experts participated. For the test, participants were presented with recordings from our target models and asked to rank the outputs based on perceived quality and intelligibility while penalizing any generative artifacts. The recordings used for the test were restored real-world samples taken from some public speeches, interviews, and call recordings. 

Iterative refinement is a widely adopted technique in image super-resolution \cite{saharia2021image}, where a low-resolution image is progressively enhanced by conditioning each super-sampling model on the output of the previous iteration. To adapt this approach for speech restoration, we applied iterative restoration using the same model across multiple iterations. This process aimed to observe how distortions introduced by the model would accumulate over repeated restoration steps. We conducted experiments with five iterations on our internal test set, evaluating the performance of each model in terms of speech quality preservation.

\begin{table}[t]
    \centering
    
    \begin{tabular}{l r r r r}
        \toprule
        \multirow{2}{*}[-2pt]{\textbf{Method}} & \multicolumn{4}{c}{\textbf{SSIM} ($\uparrow$)} \\
        \cmidrule{2-5} 
        & $16kHz$ & $24kHz$ & $44.1kHz$ & $48kHz$ \\
        \midrule
        VoiceFixer      & $0.33$ & $0.44$ & $ 0.55$ & $ 0.59$ \\
        Resemble Enhance & $0.35$ & $0.45$ & $0.57$ & $0.61$ \\
        Hi-ResLDM             & $\textbf{0.55}$ & $\textbf{0.61}$ & $\textbf{0.66}$ & $\textbf{0.65}$ \\
        \bottomrule 
    \end{tabular}
    \vspace{0.1cm}
    \caption{Structural similarity of our approach with VoiceFixer and Resemble Enhance at different sampling rates on the Valentini test-set. A higher value of SSIM represents higher perceptual similarity.}
    \label{res:tab:ssimscore}
\end{table}

\begin{table}[t]
\centering
\begin{tabular}{c c c}
\toprule
Hi-ResLDM & RE & VF \\ 
\midrule
\textbf{60.83} & 29.58  & 9.59  \\ 
\bottomrule
\end{tabular}
\caption{The results of preference test comparing Hi-ResLDM with Resemble Enhance and VoiceFixer on real-world recordings. The table reflects the preference of 20 volunteering audio experts on real-world restored outputs.}
\label{tab:subjecitve_results}
\end{table}

%% file: sections/results.tex
\section{Results}

We present the results of our evaluation in Table \ref{res:tab:overall results}, using the testing split of the Valentini dataset \cite{valentini2017noisy} - a standard for evaluation of speech enhancement methods - and our internal evaluation dataset as described in \ref{sec:speech_data_eval}. The recovery stage performs exceptionally well on the Valentini dataset due to its focus on additive noise and achieves higher intelligibility.
Despite being a generative approach, Hi-ResLDM shows strong performance on the intrusive eSTOI metric and achieves a low WER. We attribute the improvement in intrusive metrics primarily to the discriminative method employed in the recovery stage. The reduced WER emphasizes the importance of increasing the SNR before applying the generative process to enable more accurate reconstructions. Table \ref{res:tab:ssimscore} shows that Hi-ResLDM achieves high structural fidelity compared to the target signals at different sample rates.
This underscores the effectiveness of conditioning the generative restoration stage on the clean estimate produced by the recovery stage, which significantly enhances the NISQA score.

Resemble Enhance (RE) outperforms all methods on the DNSMOS metric, with Hi-ResLDM performing comparably to RE. Upon closer examination, we found that DNSMOS penalizes our restored signals specifically on the signal quality score (DNSMOS\_SIG) while providing comparable scores for background noise removal and overall quality. A likely explanation is the presence of pre-phonemic breathing sounds in our training dataset, which is prominent in the wide-band frequency range. Since DNSMOS also evaluates within a wide-band frequency range, it incorrectly interprets short breathing sounds as signal issues, leading to a penalty in the signal quality score. Although these sounds can be removed with filtering, we explicitly avoid doing so as it would potentially introduce unforeseen biases and can affect the perceived quality of the outputs.

VoiceFixer (VF) underperforms across all test data, exhibiting significantly reduced intelligibility on our internal dataset, particularly in cases of low SNR. This performance decline is largely due to hallucinations of the GAN-based model during phone regeneration. 

Another recently proposed model named Universe++ \cite{scheibler_universalpp_2024}, which improves Universe \cite{serra_universal_2022}, has demonstrated considerable potential in enhancing regeneration fidelity. However, we could unfortunately not include it in our comparison, because it was trained on the datasets that are commonly used by the community for evaluation and testing purposes, which we utilized accordingly in our experiments. Reproducing their results is not feasible due to their reliance on proprietary internal datasets for training. Additionally, the Universe model is not publicly accessible, restricting broader comparative analysis.

\begin{figure}
    \centering
    \includegraphics[width=0.7\linewidth]{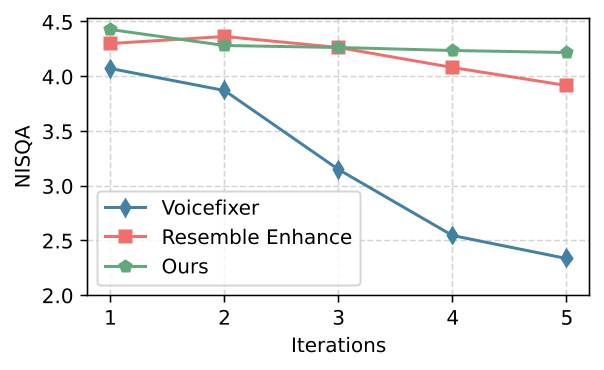}
    \caption{The figure illustrates the trend in speech quality across different speech restoration models over five iterative refinement steps.}
    \label{fig:iterative_refinement}
\end{figure}

Regarding the iterative refinement, Fig. \ref{fig:iterative_refinement} illustrates the results over five iterations, indicating that none of the speech restoration approaches improved the speech quality. However, compared to the other models, Hi-ResLDM does not substantially degrade the overall speech quality, as measured by the NISQA MOS score, indicating greater robustness than the alternatives.

\begin{figure}
    \centering
    \includegraphics[width=0.6\linewidth]{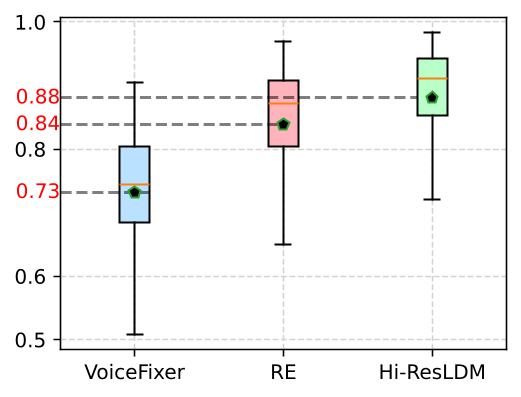}
    \caption{Plot showing the distribution of speaker recognition cosine similarity (SR-CS) of comparative models. The x-axis shows the name of the model, and the y-axis shows the cosine similarity value.}
    \label{fig:speaker_consistency}
\end{figure}

Finally, Table \ref{tab:subjecitve_results} shows the results of our subjective evaluation, where evaluators preferred the output of Hi-ResLDM in an overwhelming majority of cases. 
While the CFM-based approach performed well, VoiceFixer was penalized for the lack of intelligibility and altering the speaker's pitch and tone. We confirm these inconsistencies through SR-CS as shown in Fig. \ref{fig:speaker_consistency}.

%% file: sections/conclusion.tex
\section{Conclusion}

In this work, we present Hi-ResLDM, an improved two-stage approach to speech restoration. The recovery stage in Hi-ResLDM assists the restoration stage, enabling the generation of studio-quality speech at a sampling rate of 48 kHz. 
Hi-ResLDM outperforms SOTA models utilizing GANs and CFMs, especially on NISQA, WER, and eSTOI. 
We also investigated the iterative refinement technique for audio, which, although successful in image restoration, did not result in noticeable improvements in speech restoration. Nonetheless, Hi-ResLDM exhibited consistent generation stability over multiple iterations.
Furthermore, our model maintained high speaker consistency, and in subjective evaluations, the output was preferred in 60.83\% of cases. 